\documentclass[runningheads,a4paper]{llncs}

\usepackage{amssymb}
\usepackage{graphicx}
\usepackage{mathpartir}

\usepackage{url}

\usepackage{hyperref}
\hypersetup{
  bookmarks=true,
  colorlinks=true,
  linkcolor=black,
  citecolor=black,
}

\urldef{\mailsc}\path|something@somewhere.top|    
\newcommand{\keywords}[1]{\par\addvspace\baselineskip
\noindent\keywordname\enspace\ignorespaces#1}

\newcommand\mizarsize\footnotesize
\newcommand\mmem{~{\mathtt{in}}~}
\newcommand\mnot{{\mathtt{not}}~}
\newcommand\mand{~{\mathtt{\&}}~}
\newcommand\mor{~{\mathtt{or}}~}
\newcommand\mimp{~{\mathtt{implies}}~}
\newcommand\miff{~{\mathtt{iff}}~}
\newcommand\eltof{\hat\in}
\newcommand\mh[1]{\ulcorner #1 \urcorner}

\begin{document}

\mainmatter  %

\title{Extracting Higher-Order Goals from the Mizar Mathematical Library}

\titlerunning{Higher-Order Goals from the MML}

\author{Chad E. Brown
\and Josef Urban\thanks{This work was supported by ERC Consolidator grant nr. 649043
\textit{AI4REASON}.}%
}
\authorrunning{Chad E. Brown \and Josef Urban}
\institute{Czech Technical University in Prague}

\toctitle{Extracting Higher-Order Goals from the Mizar Mathematical Library}
\tocauthor{Chad E. Brown, Josef Urban}
\maketitle

\begin{abstract}
Certain constructs allowed in Mizar articles cannot be 
represented in first-order logic but can be represented
in higher-order logic.  We describe a way to obtain
higher-order theorem proving problems from Mizar articles that make
use of these constructs.  In particular, higher-order logic is used to
represent schemes, a global choice construct and set level binders.
The higher-order automated theorem provers Satallax and LEO-II have been run on
collections of these problems and the results are discussed.
\keywords{Formalized Mathematics, Set Theory, Higher-Order Logic, Automated Theorem Proving}
\end{abstract}

\section{Introduction}

The Mizar Problems for Theorem Proving (MPTP) system
has been developed and used
to extract first-order theorem proving problems from the Mizar Mathematical Library (MML)~\cite{Urban2003,Urban2005,Urb06}.
However, some aspects of the Mizar language cannot be directly represented in first-order.
In particular, Mizar provides supports for {\em{Schemes}} (allowing some degree of quantification over predicates and functions),
{\em{Fraenkel terms}} (allowing sets to be specified using term level binders such as $\{f(x)|x\in A, p(x)\}$)
and a {\em{global choice operator}} {\tt{the}} on types~\cite{JFR1980}.
In order to obtain first-order problems, the MPTP has dealt with schemes used in a proof
by exporting the first-order instances of the scheme used in the proof.
Additionally, Fraenkel terms and global choice have been made first-order by a process of deanonymization~\cite{Urb06}.

We describe an extension of MPTP targeting higher-order logic.
Schemes can be represented directly in higher-order logic
since quantifiers over predicates and functions are allowed.
Instead of giving the instances of schemes used in a proof, 
schemes are exported as second-order formulas (relying on the problem solver to find appropriate instances).
Global choice can be represented by a selection operator on
the type of individuals and a corresponding choice axiom.
We also give a method for representing Fraenkel terms,
though these are more challenging both to represent and to reason about.

The resulting system has been used to extract a collection of higher-order theorem proving problems
in THF0 format~\cite{SB2010}. As in~\cite{Urb06} we can partition the problem set into
simple justifications (the Mizar {\tt{by}} steps -- or sometimes no explicit justification),
scheme justifications (the Mizar {\tt{from}} steps indicating application of a scheme)
and theorems (including schemes proven in the MML).
There are roughly 10192 scheme justifications throughout Mizar proofs in the MML,
and we consider the higher-order problems corresponding to all of them.
For simple justifications, we focus only on those involving global choice or Fraenkel terms
and restrict ourselves to such steps in only four Mizar articles,
giving 245 higher-order problems involving Fraenkel terms and 47 problems involving the global choice operator.
For theorems, we focus only on 610 proven schemes whose proof in the MML requires a scheme justification.
We describe some examples and results from running
the higher-order automated theorem provers Satallax~\cite{Brown2012a} and LEO-II~\cite{LEO2015}
on some collections of these problems.

In Section~\ref{sec:hol}
we give a short description of the syntax of higher-order logic.
In Section~\ref{sec:mizar}
we define M-types, M-terms and M-propositions
corresponding to an idealized version of the Mizar language.
In Section~\ref{sec:map}
we describe the mapping of M-types, M-terms and M-propositions
into higher-order terms,
with a focus on the higher-order aspects.
Section~\ref{sec:experiments} describes experiments
using Satallax and LEO-II on the resulting higher-order problems.

\section{Syntax of Higher-Order Logic}\label{sec:hol}

We give a short introduction to the syntax of higher-order logic (in the form of Church's simple type theory~\cite{Church40})
so that we can describe the mapping in Section~\ref{sec:map}.
In order to present higher-order problems to theorem provers, the THF0 format is used~\cite{SB2010},
but we mostly restrict ourselves to mathematical presentations of higher-order terms here.

There are two base types $o$ (for propositions) and $\iota$ (for individuals, which will always be sets for us).
The remaining types are function types $\alpha\beta$ where $\alpha$ and $\beta$ are types.
The type $\alpha\beta$ is the type of functions from $\alpha$ to $\beta$ (and is sometimes written $\alpha\to\beta$).

We assume there are infinitely many variables $x$ at each type $\alpha$. We sometimes write the type as a subscript
to make it clear, as in $x_\alpha$.
Likewise, there may be arbitrarily many constants $c$ at each type $\alpha$.
We freely generate the set of typed terms as follows:
\begin{itemize}
\item A variable $x$ of type $\alpha$ is a term of type $\alpha$.
\item A constant $c$ of type $\alpha$ is a term of type $\alpha$.
\item If $s$ is a term of type $\alpha\beta$ and $t$ is a term of type $\alpha$, then $(st)$ is a term of type $\beta$.
\item If $x$ is a variable of type $\alpha$ and $s$ is a term of type $\beta$, then $(\lambda x.s)$ is a term of type $\alpha\beta$.
\item $\top$ is a term of type $o$.
\item If $s$ and $t$ are terms of type $\alpha$, then $(s=_\alpha t)$ is a term of type $o$.
\item If $s$ is a term of type $o$, then $(\neg s)$ is a term of type $o$.
\item If $s$ and $t$ are terms of type $o$, then $(s\land t)$, $(s\lor t)$, $(s\to t)$ and $(s\leftrightarrow t)$
  are terms of type $o$.
\item If $x$ is a variable of type $\alpha$ and $s$ is a term of type $o$, then $(\forall x.s)$ and $(\exists x.s)$ are terms of type $o$.
\end{itemize}
Terms of type $o$ are also called propositions.

We omit parentheses with the following conventions:
\begin{itemize}
\item Application associates to the left, e.g., $stu$ means $((st)u)$.
\item Binders have as large a scope as possible, e.g., both $x$ are bound in $\forall x_o.x\lor\neg x$.
\item The connectives $\to$, $\land$ and $\lor$ are considered right associative.
\item The precedence of the binary and unary connectives are $=_\alpha$, $\neg$, $\land$, $\lor$, $\to$ and finally $\leftrightarrow$.
\end{itemize}
In addition, we omit the type subscript on $=$ when it is clear, and
we write $s\not= t$ for $\neg (s=t)$.
Likewise we may write several binders together, as in $\forall x y z_\alpha.s$ for $\forall x.\forall y.\forall z.s$
where $x$, $y$ and $z$ should all have type $\alpha$.

\section{Idealized Mizar}\label{sec:mizar}

In order to describe the translation from Mizar to Higher-Order Logic
we first give a short presentation of an
idealized subset of the Mizar language.
For a full presentation of the Mizar language,
we direct the reader to~\cite{JFR1980}.

To simplify the presentation, we assume that some variables and constants of higher-order logic
are also variables and constants of Mizar, and that the translation will simply map
variables and constants to themselves.
The language of Mizar is restricted in a way that only variables and constants of certain
types can be used:
\begin{itemize}
\item We call variables of type $\iota$ {\emph{object variables}} and call constants of type $\iota$ {\emph{object constants}}.
\item For each $n\geq 1$, we call variables of simple type ${\underbrace{\iota\ldots\iota}_n} \iota$ {\emph{function variables (of arity $n$)}}.
Likewise, we call constants of this type {\emph{function constants (of arity $n$)}}.
We use $F$ and $G$ to range over function variables and $f$ and $g$ to range over function constants.
\item For each $n\geq 0$, we call variables of simple type ${\underbrace{\iota\ldots\iota}_n} o$ {\emph{predicate variables (of arity $n$)}}.
Likewise, we call constants of this type {\emph{predicate constants (of arity $n$)}}.
We use $P$ and $Q$ to range over predicate variables and $p$ and $q$ to range over predicate constants.
\end{itemize}
Mizar quantifiers only bind object variables. Predicate variables and function variables
only appear in schemes and are listed (with typing information) in the prefix of a scheme.

Mizar articles typically consist of definitions and theorems (some of which are schemes).
A definition may be of an object constant, a function constant or a predicate constant.
Predicate constants are sometimes defined as {\emph{modes}} or {\emph{attributes}},
which can then be used to construct Mizar types.
Mizar types can be thought of as predicates over the universe of discourse.
Mizar insists that types are nonempty and that all types, terms and propositions are well-typed (in Mizar's typing system).

In our idealized version of Mizar, we can ignore these restrictions
and define more liberal sets of
M-types, M-terms and M-propositions by mutual recursion.
The intention is that Mizar types, terms and propositions (at least within the subset of Mizar considered in this article)
will give M-types, M-terms and M-propositions, although not all M-types, M-terms and M-propositions
would be accepted by Mizar.

M-types $A,B,\ldots$ are generated as follows:
\begin{itemize}
\item ${\mathtt{set}}$ is an M-type.
\item If $p$ is an $n+1$-ary predicate constant and $T_1,\ldots T_n$ are M-terms, then $p(\cdot,T_1,\ldots,T_n)$ is an M-type.
(Here $p$ is playing the role of a Mizar {\emph{mode}}.)
\item If $q$ is a unary predicate constant and $A$ is an M-type, then $q~A$ and ${\mathtt{non}}~q~A$
are M-types.
(Here $q$ is playing the role of a Mizar {\emph{attribute}}.)
\end{itemize}
M-terms $S,T,\ldots$ are generated as follows:
\begin{itemize}
\item An object variable $x$ is an M-term.
\item An object constant $c$ is an M-term.
\item If $F$ is a function variable of arity $n$ and $T_1,\ldots T_n$ are M-terms, then\\ $F(T_1,\ldots,T_n)$ is an M-term.
\item If $f$ is a function constant of arity $n$ and $T_1,\ldots T_n$ are M-terms, then $f(T_1,\ldots,T_n)$ is an M-term.
\item If $A$ is an M-type, then $({\mathtt{the}}~A)$ is an M-term.
(The ${\mathtt{the}}$ is called a {\emph{global choice operator}}.)
\item If $x_1,\ldots,x_n$ are object variables, $A_1,\ldots,A_n$ are M-types, $T$ is an M-term
and $\Phi$ is an M-proposition, then $\{T~{\mathtt{where}}~x_1~{\mathtt{is}}~A_1,\ldots x_n~{\mathtt{is}}~A_n: \Phi\}$ is an M-term.
(These are called {\emph{Fraenkel terms}}.)
\end{itemize}
M-propositions $\Phi,\Psi,\ldots$ are generated as follows:
\begin{itemize}
\item If $P$ is an $n$-ary predicate variable of arity $n$ and $T_1,\ldots T_n$ are M-terms, then $P(T_1,\ldots,T_n)$ is an M-proposition.
\item If $p$ is an $n$-ary predicate constant of arity $n$ and $T_1,\ldots T_n$ are M-terms, then $p(T_1,\ldots,T_n)$ is an M-proposition.
\item If $S$ and $T$ are M-terms, then $(S=T)$ and $(S \mmem T)$ are M-propositions.
\item If $\Phi$ is an M-proposition, then $(\mnot \Phi)$ is an M-proposition.
\item If $\Phi$ and $\Psi$ are M-propositions, then $(\Phi\mand \Psi)$, $(\Phi\mor \Psi)$, $(\Phi\mimp \Psi)$ and $(\Phi\miff \Psi)$ are M-propositions.
\item If $x$ is an object variable, $A$ is an M-type and $\Phi$ is an M-proposition, then 
$({\mathtt{for}}~x~{\mathtt{being}}~A~{\mathtt{holds}}~\Phi)$ and 
$({\mathtt{ex}}~x~{\mathtt{being}}~A~{\mathtt{st}}~{\Phi})$
are M-propositions.
\end{itemize}

Most Mizar theorems correspond to M-propositions.
However, in some cases (namely, schemes) there are function variables or predicate variables which
cannot be bound by quantifiers.
We now define the notion of a prefix to list such variables.
When translating to higher-order propositions, the prefix will determine the outermost quantifiers.

A variable declaration is one of the following:
\begin{itemize}
\item $x:A$ where $x$ is an object variable and $A$ is an M-type.
\item $F(A_1,\ldots,A_n):B$ where $F$ is a function variable of arity $n$ and $A_1,\ldots,A_n,B$ are M-types.
\item $P[A_1,\ldots,A_n]$ where $P$ is a predicate variable of arity $n$ and $A_1,\ldots,A_n$ are M-types.
\end{itemize}
A prefix is a list of variable declarations.

An M-statement $(\Gamma,\Phi)$ is a prefix $\Gamma$ and an M-proposition $\Phi$.
For Mizar theorems other than schemes, the prefix $\Gamma$ will always be empty.
Some Mizar schemes will declare what appears to be a function variable of arity $0$.
In such a case, we use object variables instead. (This is why object variable declarations are
allowed in a prefix.)

An example of a scheme is Separation:
for each set $A$ and predicate $P$, there is a set $X$ such that $x\in X$ iff $x\in A$ and $P(x)$~\cite{XBOOLE_0.ABS}.
{\mizarsize
\begin{verbatim}
scheme Separation { A()-> set, P[set] } :
  ex X being set st for x being set holds x in X iff x in A() & P[x]
\end{verbatim}
}
\noindent
The M-statement in this case is $(\Gamma,\Phi)$ where
$\Gamma$ is the prefix $A:{\mathtt{set}},P[{\mathtt{set}}]$
(declaring an object variable $A$ of type ${\mathtt{set}}$
and a predicate variable $P$ of arity $1$) and $\Phi$ is the M-proposition
  $${\mathtt{ex}}~X~{\mathtt{being}}~{\mathtt{set}}~{\mathtt{st}}~{\mathtt{for}}~x~{\mathtt{being}}~{\mathtt{set}}~{\mathtt{holds}}~x\mmem X\miff x\mmem A \mand P(x)$$
corresponding to the body of the scheme.

\section{Mapping Mizar to Higher-Order Logic}\label{sec:map}

We now describe a mapping from M-types, M-terms, M-propositions and M-statements to higher-order terms,
concentrating on the aspects that require higher-order constructs.
The base type $\iota$ will correspond to Mizar objects (sets).
We will use $\mh{-}$ to denote the image of an M-type, M-term, M-proposition or M-statement
as a term in higher-order logic under the translation.
The intention is that mapping $\mh{-}$ should send M-statements corresponding to
Mizar theorems to provable propositions in higher-order logic.
To be precise about this would require giving details about the proof theory of Mizar,
which is beyond the scope of this paper.

In order to specify the translation, we need to declare a family of constants
the higher-order problems may make use of.
A special relation in Mizar is set membership ({\tt{in}}),
translated as {\tt{r2\_hidden}} by the MPTP system.
For this reason, we include a declared constant {\tt{r2\_hidden}}
of type $\iota\iota o$ in the higher-order setting.
For readability, we will write $s\in t$ for the term ${\mathtt{r2\_hidden}}~s~t$.
(We will also write $s\not\in t$ for $\neg(s\in t)$.)
This allows us to translate an M-proposition $S\mmem T$ simply as $\mh{S}\in\mh{T}$.
We also declare a constant $\varepsilon$ of type $(\iota o)\iota$.
This allows use to translate an M-term $({\mathtt{the}}~A)$ as $\varepsilon \mh{A}$.
Finally, we need a family of constants for translating Fraenkel terms.
For this purpose we declare a constant ${\mathtt{replSep}}_n$
of type
$$(\iota o)(\iota\iota o)\cdots ({\underbrace{\iota\cdots\iota}_{n} o}) ({\underbrace{\iota\cdots\iota}_{n} \iota}) ({\underbrace{\iota\cdots\iota}_{n} o})\iota$$
for each $n$. (In practice only a finite number of these can be declared in a single problem,
and we declare them up to the maximum $n$ required to translate the problem.
When translating the MML the maximum required $n$ was $6$.)
We can use ${\mathtt{replSep}}_n$ to translate 
$\{T~{\mathtt{where}}~x_1~{\mathtt{is}}~A_1,\ldots x_n~{\mathtt{is}}~A_n: \Phi\}$
as
{\footnotesize
$${\mathtt{replSep}_n}\mh{A_1}(\lambda x_1.\mh{A_2})\cdots(\lambda x_1\cdots x_{n-1}.\mh{A_n})(\lambda x_1\cdots x_n.\mh{T})(\lambda x_1\cdots x_n.\mh{\Phi}).$$
}

Before giving the translation, let us also remark on the intended semantics of these new constants.
The constant $\varepsilon$ is a choice operator so that $\varepsilon p$ satisfies $p$ unless $p$ is empty.
The remaining constants are set theory related, and are required since the Mizar language targets set theory.
In particular, the MML is based on Tarski-Grothendieck Set Theory (TG).
For this reason, we take the intended interpretation of $\iota$ as a model of TG.
The constant ${\mathtt{r2\_hidden}}$ is intended to be membership on this model.
The ${\mathtt{replSep}_n}$ constants give ways to specify sets. For simplicity, we consider only the $n=1$ case.
A first approximation would be to think of ${\mathtt{replSep}_1}~s~(\lambda x.t)~(\lambda x.u)$ as a set
$\{t|x\in s, u\}$. However, $s$ has type $\iota o$, not type $\iota$,
so we should write $\{t|x:sx\land u\}$.
In general, if $s$ is a predicate
that corresponds to a class instead of a set, $\{t|x:sx\land u\}$ will not be a set.
Mizar avoids this problem by enforcing an extra condition when Fraenkel terms are used:
all the types $A_1,\ldots,A_n$ must satisfy a ``sethood'' condition: that the collection of all elements of the type 
are contained in a bounding set.
In the higher-order problems we define a corresponding constant ${\mathtt{sethood}}$ of type $(\iota o)o$ as follows:
$$\lambda p_{\iota o}.\exists y_\iota .\forall x_\iota.px\to x\in y.$$
Then we can interpret ${\mathtt{replSep}_1}~s~(\lambda x.t)~(\lambda x.u)$
to be $\{t|x : sx \land u\}$ if ${\mathtt{sethood}}~s$ holds and
interpret ${\mathtt{replSep}_1}~s~(\lambda x.t)~(\lambda x.u)$
to be the empty set otherwise.
The new constants and corresponding axioms for the higher-order problems are given in Figure~\ref{fig:include}.
For each $n$ there are two axioms for ${\mathtt{replSep}}_n$: an introduction axiom ${\mathtt{replSepI}}_n$
and an elimination axiom ${\mathtt{replSepE}}_n$. The sethood conditions are only required for ${\mathtt{replSepI}}_n$
since the intended interpretation of ${\mathtt{replSep}}_n$ is the empty set when applied to an argument for which the sethood condition is violated.
In practice, ${\mathtt{sethood}}$ and ${\mathtt{replSep}}_n$ (for $n\geq 1$) are only included if the problem contains
a Fraenkel term.

\begin{figure}
{\footnotesize{
\begin{center}
\begin{mathpar}
\varepsilon : (\iota o)\iota \and {\mathtt{epsax}}:\forall p_{\iota o}.\forall x_\iota.px\to p(\varepsilon p) \and
{\mathtt{r2\_hidden}} : \iota\iota o \and
{\mathtt{sethood}} : (\iota o) o := \lambda p_{\iota o}.\exists y_\iota .\forall x_\iota.px\to x\in y\and
{\mathtt{replSep}}_1 : (\iota o)(\iota\iota)(\iota o)\iota \and
{\mathtt{replSepI}}_1 : \forall A_{\iota o}.\forall f_{\iota\iota}.\forall P_{\iota o}.\forall x_\iota.{\mathtt{sethood}}~A\to Ax\to Px\to fx\in {\mathtt{replSep}}_1~A~f~P \\
{\mathtt{replSepE}}_1 : \forall A_{\iota o}.\forall f_{\iota\iota}.\forall P_{\iota o}.\forall y_\iota.y\in {\mathtt{replSep}}_1~A~f~P \to \exists x_\iota. Ax \land Px \land y=fx \and
{\mathtt{replSep}}_2 : (\iota o)(\iota\iota o)(\iota\iota\iota)(\iota\iota o)\iota \and
{\mathtt{replSepI}}_2 : \forall A_{\iota o}.\forall B_{\iota\iota o}\forall f_{\iota\iota\iota}.\forall P_{\iota\iota o}.\forall x y_\iota.{\mathtt{sethood}}~A
  \to (\forall x_\iota.Ax\to {\mathtt{sethood}}~(Bx))
  \to\\ Ax\to Bxy\to Pxy\to fxy\in ({\mathtt{replSep}}_2~A~B~f~P) \\
\cdots \\
{\mathtt{replSep}}_n : (\iota o)(\iota\iota o)\cdots ({\underbrace{\iota\cdots\iota}_{n} o}) ({\underbrace{\iota\cdots\iota}_{n} \iota}) ({\underbrace{\iota\cdots\iota}_{n} o})\iota \\
\cdots \\
\end{mathpar}
\end{center}
}}
\caption{Higher-Order Declarations}\label{fig:include}
\end{figure}

Each M-type $A$ will map to a term $\mh{A}$ of type $\iota o$ (a predicate or class),
each M-term $T$ will map to a term $\mh{T}$ of type $\iota$ (a set)
and each M-proposition $\Phi$ will map to a term $\mh{\Phi}$ of type $o$ (a proposition).
Note that Mizar has dependent types and so
an M-type $A$ and the corresponding predicate $\mh{A}$
may contain free variables.
The mapping is defined by recursion as given in Figure~\ref{fig:trans}.
Note that while we take $\mh{x}=x$ and $\mh{c}=c$ in principle,
variables and constants are mapped to THF0 compliant names in practice.
\begin{figure}
{\footnotesize{
\begin{center}
\begin{mathpar}
\mh{\mathtt{set}} = \lambda x.\top \and
 \mh{p(\cdot,T_1,\ldots,T_n)} = \lambda x.p~x~\mh{T_1}~\ldots~\mh{T_n}^* \and
 \mh{q~A} = \lambda x.q~x\land \mh{A} x^* \and
 \mh{{\mathtt{non}}~q~A} = \lambda x.\neg q~x\land \mh{A} x^* \and
 \mh{x} = x \and
 \mh{c} = c \and
 \mh{F(T_1,\ldots,T_n)} = F~\mh{T_1}~\ldots~\mh{T_n} \and
 \mh{f(T_1,\ldots,T_n)} = f~\mh{T_1}~\ldots~\mh{T_n} \and
 \mh{{\mathtt{the}}~A} = \varepsilon \mh{A} \and
 \mh{\{T~{\mbox{where}}~x_1 {\mbox{ is }} A_1,\ldots x_n {\mbox{ is }} A_n: \Phi\}} = {\mathtt{replSep}_n}~\mh{A_1}~(\lambda x_1.A_2)\cdots~(\lambda x_1\cdots x_{n-1}.\mh{A_n})~(\lambda x_1\cdots x_n.\mh{T})~(\lambda x_1\cdots x_n.\mh{\Phi}) \and
 \mh{P(T_1,\ldots,T_n)} = P~\mh{T_1}~\ldots~\mh{T_n} \and
 \mh{p(T_1,\ldots,T_n)} = p~\mh{T_1}~\ldots~\mh{T_n} \and
 \mh{S=T} = \mh{S} =_\iota \mh{T} \and
 \mh{S \mmem T} = \mh{S} \in \mh{T} \and
 \mh{\mnot \Phi} = \neg\mh{\Phi} \and
 \mh{\Phi\mand\Psi} = \mh{\Phi\land\Psi} \and
 \mh{\Phi\mor\Psi} = \mh{\Phi\lor\Psi} \and
 \mh{\Phi\mimp\Psi} = \mh{\Phi\to\Psi} \and
 \mh{\Phi\miff\Psi} = \mh{\Phi\leftrightarrow\Psi} \and
 \mh{{\mathtt{for}}~x~{\mathtt{being}}~A~{\mathtt{holds}}~\Phi} = \forall x.\mh{A}x\to\mh{\Phi} \and
 \mh{{\mathtt{ex}}~x~{\mathtt{being}}~A~{\mathtt{st}}~\Phi} = \exists x.\mh{A}x\land\mh{\Phi}
\end{mathpar}
$^*$ where $x$ is a fresh variable of type $\iota$
\end{center}
}}
\caption{Definition of the translation}\label{fig:trans}
\end{figure}
In order to map Mizar schemes we define $\mh{(\Gamma,\Phi)}$ for M-statements by a final recursion over the prefix $\Gamma$:
\begin{itemize}
\item $\mh{(\cdot,\Phi)} = \mh{\Phi}$.
\item $\mh{((x:A,\Gamma),\Phi)} = \forall x.\mh{A}~x\to \mh{(\Gamma,\Phi)}$.
\item $\mh{((F(A_1,\ldots,A_n):B,\Gamma),\Phi)} = \forall F.(\forall x_1.\mh{A_1}~x_1\to\ldots\to\forall x_n.\mh{A_n}~x_n\to \mh{B}~(Fx_1\cdots x_n))\to \mh{(\Gamma,\Phi)}$.
\item $\mh{((P[A_1,\ldots,A_n],\Gamma),\Phi)} = \forall P.\mh{(\Gamma,\Phi)}$.
\end{itemize}

As a Mizar development is processed, new definitions are processed
and the corresponding higher-order information must be declared in the problems which use
this new information. We consider a few examples from early in the MML.

A simple example of a definition of an attribute is {\tt{empty}} given in {\tt{xboole\_0}}~\cite{XBOOLE_0.ABS}:

{\mizarsize
\begin{verbatim}
definition
  let X be set;
  attr X is empty means
  :Def1:
  not ex x being set st x in X;
end;
\end{verbatim}
}
\noindent
MPTP creates a name {\tt{v1\_xboole\_0}} of type $\iota o$.
Note that simply due to its type, {\tt{v1\_xboole\_0}} can be used as an attribute and mode to form M-types.
It can also be used to form M-propositions.
In the Mizar development, ${\mathtt{empty}}$ the proposition $X~{\mathtt{is~empty}}$
corresponds to the M-proposition ${\tt{v1\_xboole\_0}}(X)$
which translates to the higher-order proposition ${\tt{v1\_xboole\_0}}~X$.
For particular problems, MPTP also exports relevant axioms about {\tt{v1\_xboole\_0}}.
For example, its definition translates to
$\lnot \exists x.\top \land x\in X$
(or, equivalently, $\lnot \exists x.x\in X$).

The most common example of a mode used in this paper is {\tt{Element of}} from the Mizar article {\tt{subset\_1}}~\cite{SUBSET_1.ABS}:

{\mizarsize
\begin{verbatim}
definition
  let X;
  mode Element of X means :Def1:
  it in X if X is non empty otherwise it is empty;
  ...
\end{verbatim}
}
Since this is the first mode definition in the article, the corresponding
name created by MPTP is {\tt{m1\_subset\_1}},
declared to have type $\iota\iota o$.
That is, {\tt{m1\_subset\_1}} expects two arguments of type $\iota$ and yields a proposition.
The Mizar type ${\mathtt{Element~of}}~X$
corresponds to the M-type ${\mathtt{m1\_subset\_1}}(\cdot,X)$ which maps to the
term $\lambda x_\iota.{\mathtt{m1\_subset\_1}}~x~X$.
Note that the dependent Mizar type ${\mathtt{Element~of}}~X$ maps
to a term of type $\iota o$ with a free variable $X$ (making the dependency explicit).
For the sake of readability, we will write
$s\eltof t$ for
${\mathtt{m1\_subset\_1}}~s~t$.
Note that since Mizar requires all types to be nonempty, the ${\mathtt{Element~of}}$ mode is
defined so that $x\eltof X$ if and only if either $X$ is nonempty and $x\in X$
or both $X$ and $x$ are empty.
That is, if $X$ is nonempty, then $x\eltof X$ if and only if $x\in X$, as expected.
However, $x\eltof\emptyset$ if and only if $x=\emptyset$, which may be surprising when it is first encountered.

Finally, we examine examples of schemes to see how M-statements are translated in practice.

The MML includes Fraenkel's Replacement axiom scheme as an axiom of TG.
As formulated in Mizar, the scheme asserts that
for each set $A$ and each binary relation $P$ on sets,
if $P$ is functional,
then there is a set $X$ such that $x\in X$ iff there is a $y\in A$ such that $P(y,x)$~\cite{TARSKI.ABS}.
In Mizar's syntax, the scheme is specified as follows:

{\mizarsize
\begin{verbatim}
scheme Fraenkel { A()-> set, P[set, set] }:
 ex X st for x holds x in X iff ex y st y in A() & P[y,x]
 provided for x,y,z st P[x,y] & P[x,z] holds y = z
\end{verbatim}
}

This can be seen as an M-statement with prefix $A:{\mathtt{set}},P[{\mathtt{set}},{\mathtt{set}}]$
and an M-proposition corresponding to the body.
The M-statement translates to the higher-order proposition
$$\forall A_\iota.\forall P_{\iota\iota o}.(\forall xyz_\iota.Pxy\land Pxz\to y=z)\to\exists X_\iota.\forall x_\iota.x\in X\leftrightarrow\exists y.y\in A \land Pyx.$$

An early application of the Fraenkel scheme is to prove Zermelo's Separation scheme discussed at the end of Section~\ref{sec:mizar},
where the corresponding M-statement is given.
The M-statement translates to the following higher-order proposition:
$$\forall A_\iota.\forall P_{\iota o}.\exists X_\iota.\forall x_\iota.x\in X\leftrightarrow x\in A\land Px.$$

For each scheme proven in the MML, the MPTP system has generated a corresponding higher-order problem in THF0 format~\cite{SB2010}.
For example, the problem corresponding to the separation scheme is {\tt{s1\_xboole\_0}}.
In order to prove {\tt{s1\_xboole\_0}} automatically, a prover would need to synthesize the
appropriate relation to use with Replacement, e.g., $\lambda x y_\iota.x=y\land Py$
where $P$ is the predicate from Separation.
At the moment, neither Satallax nor LEO-II can prove this automatically.

The Mizar proof begins by defining a predicate $Q$ and then applying Replacement with $Q$.

{\mizarsize
\begin{verbatim}
  defpred Q[set,set] means $1 = $2 & P[$2];
A1: for x,y,z st Q[x,y] & Q[x,z] holds y = z;
  consider X such that
A2: for x holds x in X iff ex y st y in A() & Q[y,x]
                                                  from TARSKI:sch 1(A1);
\end{verbatim}
}
In $\lambda$-notation, the definition of $Q$ is $\lambda xy_\iota.x=y\land Py$.
Line {\tt{A1}} justifies $y=z$ whenever $Qxy$ and $Qxz$.
When schemes are used to justify Mizar proof steps, the keyword {\tt{from}} is used.
These are the steps we classify as {\emph{scheme justifications}}.
In this case, the Replacement scheme is used to justify the existence of a set $X$
such that $x\in X$ iff $\exists y.y\in A\land Qyx$.
A higher-order problem can be extracted from each such scheme justification.
For this particular example, the conjecture to prove is
$\exists X.\forall x.x\in X\leftrightarrow \exists y.y\in A\land y=x \land Px$.
This follows from Replacement and {\tt{A1}}, but requires instantiating the higher-order variable
in the Replacement axiom with $Q$.
Note that $Q$ is not explicitly given in the problem, but can easily be
recovered using pattern unification~\cite{MiNa87}, as we now demonstrate.
Suppose we replace the outermost quantifiers in the Replacement axiom with
existential variables
${\mathcal{A}}$ of type $\iota$ and ${\mathcal{R}}$ of type $\iota\iota o$.
The conclusion of the implication has the following form:
$$\exists X_\iota.\forall x_\iota.x\in X\leftrightarrow\exists y.y\in {\mathcal{A}} \land {\mathcal{R}}yx.$$
Since the subterm ${\mathcal{R}}yx$ is the higher-order existential variable ${\mathcal{R}}$
applied to distinct bound variables ($y$ and $x$),
we can use pattern unification (in this case pattern matching) to obtain solutions for ${\mathcal{A}}$ and ${\mathcal{R}}$.
That is, when we match against
$$\exists X.\forall x.x\in X\leftrightarrow \exists y.y\in A\land y=x \land Px$$
we obtain the disagreement pairs
$X,x,y|{\mathcal{A}} =^? A$ and
$X,x,y|{\mathcal{R}}yx =^? y=x \land Px$
which has the unique (desired) solution: $A$ for ${\mathcal{A}}$ and $\lambda yx.y=x \land Px$ for ${\mathcal{R}}$.
Neither Satallax nor LEO-II re-prove this scheme justification within 5 minutes
with the default strategy schedule.
However, Satallax is able to prove the problem corresponding to this scheme justification
under certain flag settings that encourage pattern unification.

\section{Experiments}\label{sec:experiments}

We now report on the results of running two higher-order automated theorem provers (Satallax and LEO-II)
on some of the problems resulting from the translation described in the previous section.
We consider four problem sets:\footnote{The THF versions of the problems discussed here are available from {\url{http://147.32.69.25/~chad/mptp_thf.tgz}}}
\begin{itemize}
\item {\bf{SimpGC}}: Simple justifications where the conclusion includes a global choice operator.
  From four Mizar articles~\cite{SUBSET_1.ABS,DOMAIN_1.ABS,GROUP_2.ABS,TOPGEN_2.ABS} $47$ problems were extracted.
\item {\bf{SimpFr}}: Simple justifications where the problem contains a Fraenkel term.
  We consider $245$ such problems arising from three Mizar articles~\cite{DOMAIN_1.ABS,GROUP_2.ABS,TOPGEN_2.ABS}.
  Since these proved to be surprisingly difficult, we also considered ``pruned'' versions of the problems
  in which the first-order theorem prover E~\cite{Schulz:AICOM-2002} indicated which axioms it used to find a corresponding first-order proof.
\item {\bf{SchJust}}: For each scheme justifications (using {\tt{from}}) in a Mizar proof in the MML, a corresponding problem was created.
  There are 10192 such problems.
\item {\bf{SchPfs}}: Out of 787 schemes proven in the MML, 610 have a proof making use of a scheme justification. For each of these 610 we
  have created a corresponding problem. Note that solving these problems requires finding a full proof, not justifying a single Mizar step in a proof.
  Hence these should be harder than the previous problem sets.
\end{itemize}
The results of running Satallax and LEO-II on the problem sets
with the default settings and a time limit of 5 minutes are shown in Table~\ref{tab:res}.
In addition, we note the number of problems both provers solved.
For the remainder of the section, we discuss the results and describe some concrete examples.

\begin{table}
  \begin{center}
  \begin{tabular}{r|c|c|c|c}
      & Total Problems & Satallax & LEO-II & Either \\\hline
    {\bf{SimpGC}} & 47 & 24 (51\%) & 28 (60\%) & 30 (64\%) \\
    {\bf{SimpFr}} & 245 & 126 (52\%) & 88 (36\%) & 165 (67\%) \\
    {\bf{SimpFr}} pruned & 245 & 159 (65\%) & 155 (63\%) & 192 (78\%) \\
    {\bf{SchJust}} & 10192 & 5608 (55\%) & 1524 (15\%) & 6072 (60\%) \\
    {\bf{SchPfs}} & 610 & 31 (5\%) & 67 (11\%) & 81 (13\%) \\
  \end{tabular}
  \end{center}
  \caption{Results on Problem Sets with 5 Minute Time Limit}\label{tab:res}
\end{table}

One of the first uses of the global choice operator in Mizar is to define a (first-order) choice
operator on sets called {\tt{choose}}~\cite{SUBSET_1.ABS}.

{\mizarsize
\begin{verbatim}
definition
  let S be set;
  func choose S -> Element of S equals
  the Element of S;
  correctness;
end;
\end{verbatim}
}
Note that no proof is given for correctness, as Mizar recognizes
that {\tt{the Element of $S$}} has type {\tt{Element of $S$}}.
Let us consider the corresponding higher-order simple justification problem.
The higher-order problem would include the declaration of $\varepsilon$ from Figure~\ref{fig:include}.
In addition, the fact that types of the form {\tt{Element of $A$}} are nonempty
is given: $\forall A_\iota.\exists B_\iota.B\eltof A$.
The conjecture to justify is $$\varepsilon (\lambda A_\iota.A\eltof c)\eltof c$$ for a fixed $c$.
This, of course, follows immediately from the two axioms and both Satallax and LEO-II can easily re-prove this simple justification.

Note that simply because a simple justification has a conclusion with a global choice operator
does not mean that the choice axiom plays a role in the justification.
Indeed, for the two examples from the problem set {\bf{SimpGC}} Satallax proves but LEO-II does not,
the proofs Satallax finds do not use the axiom about $\varepsilon$.
Furthermore, upon inspection it became clear that some problems neither prover could solve also do not require
the axiom about $\varepsilon$. Consider the following fragment of a Mizar proof about group theory~\cite{GROUP_2.ABS}.

{\mizarsize{
\begin{verbatim}
    set a = the Element of G;
    ...
    consider b such that
A4: H * a = {b} by A1;
    h * a in H * a by A3,Th104;
    then
A5: h * a = b by A4,TARSKI:def 1;
\end{verbatim}
}}
\noindent
The final justification is essentially the definition of singleton. The only reason the corresponding
higher-order problem falls into class {\bf{SimpGC}} is because $a$ is $\varepsilon (\lambda x.x\eltof (c~G))$
(where $c$ is a function taking a group to its carrier set, left implicit in the Mizar text).
The fact that neither Satallax nor LEO-II could solve this problem was due to the fact that
there are too many extra (unnecessary) axioms given in the generated problem.
After pruning away the unnecessary axioms (with the help of E prover on a corresponding first-order problem),
both Satallax and LEO-II can prove the pruned problem. LEO-II proves the pruned problem within 8 seconds
and Satallax proves the pruned problem in less than a second.

We now turn to the problem set ${\bf{SimpFr}}$: simple justifications involving at least one Fraenkel term,
either in the conclusion or in one of the assumptions MPTP included in the problem.
There were 640 such examples in the four Mizar articles we considered,
but with experimentation it became clear that often the Fraenkel term was in an assumption
that was unnecessary for the proof.
In order to obtain a reasonable problem set, we used E on corresponding first-order problems to obtain
pruned versions of the 640 problems. (In cases where E could not find the proof, we omitted the problem.)
After pruning, there were 245 problems that still included a Fraenkel term.
On each of these 245 problems, we ran Satallax and LEO-II on both the original and pruned problems.
On the original versions, only 20\% of the problems could be solved by both provers,
whereas on the pruned versions, 50\% could be solved by both provers.
This suggests that better relevance filtering would be one of the most important potential improvements.

We briefly examine two small examples involving Fraenkel terms.
Consider the following proof fragment from~\cite{DOMAIN_1.ABS}. %

{\mizarsize
\begin{verbatim}
assume a in { x1 : x1 in A1 & not x1 in B1 or not x1 in A1 & x1 in B1 };
then ex x1 st a = x1 &
                   (x1 in A1 & not x1 in B1 or not x1 in A1 & x1 in B1);
\end{verbatim}
}
\noindent
In the context of this fragment, {\tt{x1}} ranges over elements of a nonempty set {\tt{X1}}.
Mizar is able to verify the correctness of the last line from
the first line without any explicit references
as this is simply the property of membership in a Fraenkel term.
In the corresponding higher-order problem, the elimination principle ${\mathtt{replSepE}}_1$
is required for the justification.
Satallax can prove the corresponding problem in less than a second.
The first mode in the default strategy schedule that finds the proof is one
making use of pattern unification.
In particular, after replacing the outermost quantifiers of
${\mathtt{replSepE}}_1$
with existential variables ${\mathcal{A}}$, ${\mathcal{F}}$, ${\mathcal{P}}$ and ${\mathcal{Y}}$,
the proposition has the form:
$${\mathcal{Y}}\in {\mathtt{replSep}}_1~{\mathcal{A}}~{\mathcal{F}}~{\mathcal{P}} \to \exists x_\iota. {\mathcal{A}}x \land {\mathcal{P}}x \land {\mathcal{Y}}={\mathcal{F}}x.$$
All the occurrences of the existential variables are pattern occurrences, and so pattern matching can be used to find
the appropriate instances.
In particular, one axiom of the problem is
$$a\in {\mathtt{replSep}}_1~(\lambda x_\iota.x\eltof X)~(\lambda x_\iota.x)~(\lambda x_\iota.x\in A\land x\notin B\lor x\notin A\land x\in B).$$
When the antecedent of the implication above is matched against this axiom, the following instantiations result:
\begin{itemize}
\item ${\mathcal{Y}} := a$
\item ${\mathcal{A}} := \lambda x_\iota.x\eltof X$
\item ${\mathcal{F}} := \lambda x_\iota.x$
\item ${\mathcal{P}} := \lambda x_\iota.x\in A\land x\notin B\lor x\notin A\land x\in B$
\end{itemize}
Given these instantiations, the solution is immediate.
Satallax can prove both the pruned and unpruned version of this example in less than a second.
LEO-II timed out after five minutes on both versions.

We consider a simple justification requiring the ${\mathtt{replSepI}}_1$. %
Consider the following proof fragment from~\cite{DOMAIN_1.ABS}:

{\mizarsize
\begin{verbatim}
A2: a = x1 and
A3: P[x1];
  Q[x1] by A1,A3;
  hence thesis by A2;
\end{verbatim}}
\noindent
where the thesis in the last step is

{\mizarsize
\begin{verbatim}
  a in { z1 where z1 is Element of X1: Q[z1] }
\end{verbatim}}
\noindent
As in the previous example, {\tt{x1}} ranges over elements of a nonempty set {\tt{X1}}.
In the higher-order problem corresponding to the final simple justification ({\tt{by A2}}), the conjecture has
the form
$a\in {\mathtt{replSep}}_1~(\lambda x_\iota.x\eltof X)~(\lambda x_\iota.x)~(\lambda x_\iota.Qx)$.
In addition ${\mathtt{replSepI}}_1$,
the axioms needed for the proof
are $x_1\eltof X$ (using the type of {\tt{x1}} in the Mizar article),
$a=x_1$ (from {\tt{A2}} in the proof fragment above), $Qx_1$ (from the previous step in the proof fragment above)
and the extra axiom $\forall X_\iota.{\mathtt{sethood}}~(\lambda x_\iota.x\eltof X)$.
Satallax requires roughly $6$ seconds before reaching a mode in the default strategy schedule that
can solve this problem. The successful mode requires less than a second to find the proof.
Again, the mode makes use of pattern unification to find the proper instantiations.
LEO-II can also find the proof in this example, and takes just under $6$ seconds.

Lastly we turn to scheme justifications ({\bf{SchJust}}) and full proofs of schemes ({\bf{SchPfs}}).
In Section~\ref{sec:map} we have already discussed an example of a scheme
that cannot be automatically proven (Separation from Replacement) by either prover.
In addition we saw that neither prover could even re-prove the relevant scheme justification
in the Mizar proof of Separation from Replacement
within 5 minutes using the default settings.

Satallax performed significantly better than LEO-II on scheme justifications,
while LEO-II performed significantly better than Satallax on proofs of full schemes.
We consider one example of a scheme justification that Satallax solved but LEO-II did not.
We then consider an example of a full scheme that LEO-II solved but Satallax did not.

The set operation $X\setminus Y$ is defined  %
in an early Mizar article~\cite{XBOOLE_0.ABS},
and the following required existence proof is given:

{\mizarsize
\begin{verbatim}
defpred P[set] means not $1 in Y;
thus ex Z being set st for x holds x in Z iff x in X & P[x]
                                                       from Separation;
\end{verbatim}
}
\noindent
Note that the scheme justification makes use of the Separation scheme
using the set $X$ and the predicate $\lambda x.x\not\in Y$.
Again, the higher-order instantiation $\lambda x.x\not\in Y$
can be determined using pattern matching,
and Satallax can re-prove this in a fraction of a second using such a mode.
With the default strategy schedule, Satallax tries such a mode and solves
the problem in $37$ seconds.
LEO-II times out after 5 minutes.

A scheme LEO-II can fully prove but Satallax cannot is the following Mizar scheme~\cite{SUBSET_1.ABS}:

{\mizarsize
\begin{verbatim}
scheme SubsetEx { A() -> non empty set, P[set] } :
  ex B being Subset of A() st
  for x being Element of A() holds x in B iff P[x]
\end{verbatim}
}
\noindent
This is again a form of Separation and is proven using the Separation scheme already considered.
The primary difference between the schemes is that the new scheme {\tt{SubsetEx}}
asserts that the set has type ${\mathtt{Subset~of}}~A$ (notation for ${\mathtt{Element~of}}~\wp A$)
and restricts the inner universal quantifier to ${\mathtt{Element~of~}}A$.
In the corresponding higher-order problem, we must prove the formula
$$\exists B. B\eltof \wp A\land \forall x.x\eltof A\to (x\in B\leftrightarrow Px)$$
from the higher-order formula
$$\forall Q_\iota\forall X_\iota.\exists B.\forall x.x\in B\leftrightarrow x\in X\land Qx.$$
The solution is simple: instantiate the assumption with the $Q:=P$ and $X:=A$
giving an appropriate witness $B$ for the conjecture.
Some minor first-order reasoning completes the proof.
LEO-II can find the proof by doing some clause normalization and calling E.
It is E that does the ``higher-order'' instantiation of $P$ for $Q$ and completes the proof.
This is possible since the higher-order problem, after being encoded into first-order, is still
provable. (In particular, the proof does not require $\beta$-reductions.)
Satallax, on the other hand, does not solve the problem and times out after 5 minutes.
The minor structural differences between the assumption and conclusion prevents
pattern matching from suggesting the instantiation $P$ for $Q$.
While $P$ is among the possible instantiations considered for $Q$,
other possible instantiations are considered as well. The combination
of multiple possible instantiations and required first-order reasoning
makes the problem out of reach for the current version of Satallax.

\section{Conclusion}

We have described an extension of MPTP that creates higher-order theorem proving problems
from the MML.
The resulting problems seem to present challenges for higher-order theorem provers.
For example, even some of the easiest problems become difficult if there are too many axioms,
so better relevance filtering is necessary.
Even simple reasoning about Fraenkel terms seems to be more difficult than one would expect,
and so these examples may provide insights into improvements that can be made
to automated provers.

There are multiple possibilities for the translation of Fraenkel terms that bind more than one set variable.
We have implemented one way and suggested another.
Further experimentation will likely be helpful for determining a good way to handle these cases.

The problems generated from scheme justifications and full proofs of schemes
turned out to show the different strengths and weaknesses of Satallax and LEO-II.
Hopefully such problem sets will lead to improvements in higher-order automated theorem provers.
Given enough improvement on such problems, perhaps higher-order automated provers
could provide help to Mizar authors who make use of the features of Mizar that go beyond first-order.
In order to serve this purpose, care would have to be taken that the automated provers do
not search for proofs that go beyond Mizar's logic (e.g., make use of higher-order quantifiers within instantiations).
We leave such concerns to future work.

\bibliographystyle{splncs03}
\bibliography{refs,fm}

\end{document}